\documentstyle[11pt,fleqn]{article}
\def\ref{par\noindent\hangindent=6mm\hangafter=1}
\begin{document}
\vbox{
\rightline{Mod. Phys. Lett. A 8 (1993) 3429-3434}
}
\baselineskip 8mm

\begin{center}
{{\bf Black Holes and Radiometry}}

\bigskip

H.C. Rosu\footnote{Electronic mail:
rosu@ifug.ugto.mx}

{\it Instituto de F\'{\i}sica de la Universidad de Guanajuato, Apdo Postal
E-143, Le\'on, Gto, M\'exico}

\end{center}

\bigskip
\bigskip

\begin{abstract}

Following Grischuk and Sidorov [Phys. Rev. D 42 (1990) 3413] 
in putting the Bogolubov-Hawking
coefficient of Schwarzschild black-holes in the squeezing perspective,
we provide a short discussion of Schwarzschild black holes as radiometric
standards.

\end{abstract}

\vskip 1cm



In this work I will give arguments in favour of taking Schwarzschild
black holes (SBH) as blackbody simulators, most probably the best ones.
Even before the discovery of their ``horizon" radiation, \cite{1},
 SBHs might
have been considered in the special class of ``material" bodies in
 the Universe to
be selected as blackbody simulators. In classical physics/relativity
the horizon surface of a SBH absorbs all the radiation falling on it.
This is nothing else but the common definition of a blackbody (i.e.,
absorptivity $\alpha=1$). The radiometric considerations to follow
stem from the connections between the theory of ``particle creation"
in external fields and the quantum-mechanical theory of squeezed
states. In the seminal paper \cite{2}, Grishchuk and Sidorov hinted
on the close relationships of the two research fields, and they
 suggested
the observed large-scale structure of the Universe to be just a
strongly squeezed state of the zero-point quantum fluctuations of
a cosmological scalar field.


In the case of a scalar quantised field in the SBH gravitational field,
a two-mode squeeze operator comes into play in order to relate the in-
 and out-vacuum. This is so because Hawking radiation is a manifestation
of the decomposition of the field over travelling waves in a region with
rather complicated causal structure, involving conditions on base fields
not only at spatial infinity, but also at the horizon. Hawking has
masterly solved this problem in 1976, \cite{3}. First he used a
decomposition of the field in the form:
$$\phi=\int d\omega (a^{(1)} _{\omega} f^{(1)} _{\omega}+
a^{(3)} _{\omega}
 f^{(3)} _{\omega} +a^{(4)} _{\omega} f^{(4)} _{\omega} + h.c.).  \eqno(1)$$
The base functions satisfy the following boundary conditions:
\[ f^{(1)} _{\omega} \approx
\left\{ \begin{array}{ll}
          e^{i\omega v} & \mbox{on $\cal I ^{-}$}\\
          0             & \mbox{on $\cal H ^{-}$}
        \end{array}
\right. \]
\[f^{(3)} _{\omega} \approx
\left \{ \begin{array}{ll}
           0                 & \mbox{on $\cal I ^{-}$}\\
           e^{i\omega u_{+}} & \mbox{on $\cal H ^{-}$}
         \end{array}
\right. \]
\[f^{(4)} _{\omega} \approx
\left \{ \begin{array}{ll}
           0                  & \mbox{on $\cal I ^{-}$}\\
           e^{-i\omega u_{+}} & \mbox{on $\cal H ^{-}$}
         \end{array}
\right. \]

The retarded and advanced variables are related to the Schwarzschild
coordinates (t,r) in the well-known way:
$$v=t+r+2M\ln (r/2M-1); U=\mp 4Me^{-u/4M},  \eqno(2)$$
\[u_{+} = \left\{ \begin{array}{ll}
                    -\kappa^{-1} \ln(-U)   & \mbox{for $U<0$}\\
                    -\kappa^{-1} \ln(U)    & \mbox{for $U>0$}
                  \end{array}
          \right. \]

$\kappa =(4GM)^{-1}$ is the surface/horizon gravity of the black hole.
The in-vacuum state is annihilated by all the annihilation operators
simultaneously:
$$a^{(1)} _{\omega} |0_{-} \rangle=a^{(3)} _{\omega} |0_{-}\rangle =
a^{(4)} _{\omega} |0_{-} \rangle    =0.    \eqno(3)$$
The same scalar field can be expanded over another set of base functions
whose annihilation operators are defined in terms of the out-vacuum,
$$\phi =\int d\omega (g_{\omega} w_{\omega} +h_{\omega} y_{\omega}
+j_{\omega} z_{\omega} +h.c.).  \eqno(4) $$
The set ({\it w,y,z}) has the following behaviour:
\[w_{\omega} \approx \left\{ \begin{array}{ll}
              0       &\mbox{on $\cal I ^{-}$}\\
              0       &\mbox{on $\cal H ^{-}$ ;$ U<0$}\\
              e^{-i\omega u_{+}}  &\mbox{on $\cal H ^{-}$ ; $U>0$}
          \end{array}
          \right. \]
\[y_{\omega} \approx \left\{ \begin{array}{ll}
             0                   &\mbox{on $\cal I ^{-}$}\\
             e^{i\omega u_{+}}   &\mbox{on $\cal H ^{-}$; $U<0$}\\
             0                   &\mbox{on $\cal H ^{-}$ ;$ U>0$}
          \end{array}
          \right. \]
\[z_{\omega} \approx \left\{  \begin{array}{ll}
             e^{i\omega u_{+}}  &\mbox{on $\cal I ^{-}$}\\
             0                  &\mbox{on $\cal H ^{-}$}
           \end{array}
           \right.  \]
The out-vacuum is the state defined by :
$$g_{\omega} |0_{+}\rangle = h_{\omega} |0_{+}\rangle =j_{\omega}
 |0_{+}\rangle =0.  \eqno(5)$$
 Hawking has obtained the following Bogolubov transformations for SBH,
 \[  \left\{  \begin{array}{ll}
       a^{(1)} _{\omega}=j_{\omega}\\
       a^{(3)} _{\omega} =(1-x_{b})^{-1/2}
 (h_{\omega} - x^{1/2} _{b} g^{\dag} _{\omega})\\
       a^{(4)} _{\omega} =(1-x_{b})^{-1/2}
  (g_{\omega} -x^{1/2} _{b} h^{\dag} _{\omega})
              \end{array}
              \right.   \]
 There is only one Bogolubov parameter $x_{b} =exp (-8\pi GM\omega)$.
  As a consequence
 of these in-out transformations, the two-mode SBH squeeze operator is
 found to be $S( r, \pi)$,
 where the squeezing parameter is related to the Bogolubov one by:
 $\tanh ^{2}r=x_{b}$.
 Remote observers, on Earth and likewise, have access to the {\it y}
particles only, and should average over the unobservable {\it w}
particles.
   One will find out a pure thermal density matrix of the form:
 $$\rho_{SBH} =(1-x_{b})\sum_{m=0}^{\infty} (x_{b})^{m} |m_{w}\rangle
  \langle m_{w} |.  \eqno(6)$$
  The most important feature of SBH problem to be emphasized here
  (which one will encounter only in
  a few other cases, like Rindler motion and de Sitter space-time)
   is that
  the density matrix is thermal in each mode, with one and the same
  universal ``thermodynamic" temperature in each mode defined as:
  $$e^{-\omega/T_{\omega}} =x_{b}.  \eqno(7)$$
  In the SBH case the modal temperature is:
  $ T_{\omega} = T_{H} = (8\pi GM)^{-1} $, a fact we believe to be of
great radiometric relevance. 


In the Rindler problem, we first remark that the quantum correlations
between the causally disconnected regions $R$ and $L$ (right and left
wedges) are of EPR type in an almost manifest way. When the observer
belonging to one of the edges is measuring his particles, he has to
trace out the states in the other wedge and he will find out a wedge
modal density matrix of the exact thermal type (cf. Eq.(6)), with the modal
temperature directly proportional to the Rindler acceleration.


  In order to better understand possible radiometric consequences of the
Schwarzschild problem, it is worth pointing
 out the relationship to the quantum-mechanical tunneling problem of the
 corresponding Schr\"{o}dinger equation for the scalar field. In this
 case one could show that $R_{\omega} = e^{-\omega /T_{H}} $
 in each mode, where $ R_{\omega}$ is the ``above-barrier" reflection
 coefficient of the mode $\omega$. In a semi-logarithmic plot
  ($\ln R_{\omega}$, $\omega$), one should find out a straight line
  with the negative slope $1/T_{H}$. We have here a quantum tunneling
  radiometric feature similar to that of the squeezing framework.

  If instead of that, one will make
  use of a scattering picture, there could be turning points in the
  potential barrier of the corresponding Schr\"{o}dinger equation.
  Whence a WKB tunneling picture will predict a different reflection
  coefficient, which is missing the very simple radiometric
  character of the squeezing picture. The scattering and absorption
  of scalar waves by Schwarzschild black holes have been considered
  by Matzner long ago, \cite{4}. On the other hand,
 Fabbri, \cite{5}, dealt with
  the more realistic case of electromagnetic waves, and studied in
  detail the tunneling through the one-dimensional barrier
  occuring for each partial wave of the modified Debye potentials
  introduced by Mo and Papas, \cite{6}. The barrier has the form:
$$U_{l} (r^{*})=\left(1-\frac{2M}{r}\right) \frac{l(l+1)}{r^{2}}, \eqno(8)$$
where $r^{*}$ and r are connected in the well known manner:
$$r^{*}=r+2M\ln \left(\frac{r}{2M}-1\right). \eqno(9)$$
The square root of the coefficient of the nonderivative term in
the one dimensional Schr\"{o}dinger equation can always be interpreted
as a local wave number, depending in the Schwarzschild case on the radial
coordinate as follows
$$K_{l} (r^{*})= \Big[\omega ^{2} - \left(1-\frac{2M}{r}\right)
 \frac{l(l+1)}{r^{2}}\Big]^{1/2}.
\eqno(10)$$
 If the potential peak is smaller than $\omega ^{2}$, this wave
 number is real with a minimum at $r=3M$, but otherwise it turns
 imaginary for the region between the two turning points given by
 $K_{l} (r^{*})=0$. For the modified Debye partial waves,
 Fabbri found the following location of the turning points:
$$r_{1}=\frac{2}{\omega}\Big [\frac{l(l+1)}{3}\Big]^{1/2} \cos\frac{\eta}{3}
\eqno(11)$$

$$r_{2}=\frac{2}{\omega} \Big[\frac{l(l+1)}{3}\Big] \cos\frac{\eta -2\pi}{3},
 \eqno(12)$$
where:
$$\eta =\arccos\Big[-3\omega M\left(\frac{3}{l(l+1)}\right)^{1/2}\Big]  
\eqno(13)$$
is the first quadrant value of the inverse trigonometric function,
implying the following order $2M\le r_{1}\le r_{2}$.

The two turning points exist in all partial waves as soon as $\omega$
is smaller than a critical angular frequency given by:
$$\omega_{c}=\left(\frac{2}{27}\right)^{1/2} \frac{1}{M}.  \eqno(14)$$
Beyond the critical frequency, the existence of turning points is
limited only to $l$ greater than a critical value given by:
$$l_{c}(l_{c}+1)=27\omega ^{2} M^{2}.  \eqno(15)$$

The reflection coefficient depends strongly on the presence or absence
of the turning points. For example when they exist, the reflection
coefficient may be written in the following analytical form:
$$ R^{2} _{l} =\frac{\exp (2\theta _{l})}{1+\exp (2\theta _{l})},
\eqno(16)$$
where $\theta_{l}$ is a complicated expression in terms of elliptical
integrals and therefore the simple features of the squeezing reflection 
coefficient do not show up in the WKB tunneling.


We would like to refer now to the thermal emission phenomena
 originating in
WKB tunneling with two turning points in the context of the field
 electron emission from metal surfaces, \cite{7}.
 The direction of the barrier penetration
is reversed as compared to the black hole case, and we have one
 cartesian coordinate, normal to the metal surface, and not the radial
 coordinate of the black hole problem. Also the shape of the barrier
 is different, and one is dealing with electrons.
  Nevertheless, we consider the
 example very instructive. The surface barrier for the electron is:
$$ V(z)=\left\{ \begin{array}{ll}
                  E_{F}+\phi-\frac{e^{2}}{4z}-eFz &
                   \mbox{for  $\;z>z_{c}$} \\
                  0       &\mbox{for  $\;z<z_{c}$}
                \end{array}
                  \right.       \eqno(17)$$
where $E_{F}$ is the Fermi energy, $\phi$ is the work function, and the
last two terms correspond to the image force and the electric field F
applied to the surface. The distance $z_{c}$ is determined by
$V(z_{c})=0$. A WKB-type approximation leads to very accurate results
for the emitted current density in the field emission region, i.e.,
emission at very low temperatures and strong applied field.
 In particular, the WKB
method reproduces the well-known Fowler-Nordheim formula for the
field emission. This is the plot of $\ln (J/F^{2})$ versus 1/F,
which is a straight line with a negative slope. For more details,
I recommend the very clear exposition in the book of Modinos,
 \cite{7}. The other regime, of low field and high temperature,
 which is known as the thermionic emission, could also be reproduced
 by the WKB method with two turning points only. This time the plot of
 $\ln J$ against $\sqrt{F}$ is a straight line with the negative slope
 $m_{S}=e^{3/2}/kT$. Such lines are called Schottky lines. We can see
 on these examples that
 the emission regimes of metal surfaces, despite differences of
 treatment and concepts, display a certain resemblance to the black hole
 plot that I have mentioned before.

\section*{ Acknowledgements}

This work was partially supported by CONACyT Grant No. F246-E9207.
The author is grateful to O. Obreg\'on for discussions and
encouragements.

The author would like to thank A. Salam,
the International Atomic
Energy Agency and UNESCO for hospitality at the International
Centre for Theoretical Physics, Trieste, where the work has been started.

\bigskip

{\bf Note added:}

In the physical optics context,
Yurke and Potasek \cite{yp} have shown that the parametric interactions
resulting in the two-mode optical squeezing provide a mechanism for
thermalization, so that if one has access to only one mode of a two-mode
squeezed vacuum, the photon statistics is indistinguishable from that of
a thermal distribution.

In the simple model of parametric down-conversion in which the signal and
idler modes are single modes, the state generated is
$\sum _{n=0}^{\infty} c_{n}|n\rangle |n\rangle$, where
$c_{n}=(-i\tanh r)^{n}/\cosh r$, and $r$ being proportional to the
parametric coupling constant and the interaction time \cite{rm}.


\end{document}